\documentclass[fleqn,twoside]{article}
\usepackage{espcrc2}

\usepackage{graphicx}
\usepackage[figuresright]{rotating}

\newcommand{\AmS}{{\protect\the\textfont2
  A\kern-.1667em\lower.5ex\hbox{M}\kern-.125emS}}

\newcommand{\bea}{\begin{eqnarray}}
\newcommand{\eea}{\end{eqnarray}}

\newcommand{\as}{\alpha_s}
\newcommand{\asMZ}{\alpha_s(M^2_Z)}

\title{The value of QCD coupling constant and power 
corrections 
in the structure function F2 measurements}

\author{V.G. Krivokhijine\address[MCSD]{
Joint Institute for Nuclear Physics,
141980 Dubna, Russia},
        A.V. Kotikov\addressmark
}

\begin{document}

\begin{abstract}
We reanalyze deep inelastic scattering data 
of BCDMS Collaboration by including proper cuts of  ranges
with large systematic errors. 
We perform also the fits of high statistic deep inelastic scattering data 
of BCDMS, SLAC, NM and BFP Collaborations  
taking the data separately and in combined way and find good agreement
between these analyses. We extract the values of both
the QCD coupling constant $\alpha_s(M^2_Z)$ up to
NLO level and of the power corrections to the structure function $F_2$. 
\vspace{1pc}
\end{abstract}

\maketitle

%\vspace{-4mm}
%\vskip -0.3cm
\section{ Introduction }

The deep inelastic scattering (DIS) leptons on hadrons is the basical
 process to study the values of the parton distribution functions 
which are universal (after choosing of factorization and renormalization 
schemes) and
can be used in other processes.
The accuracy of the present data for deep inelastic
structure functions (SF) reached the level at which
the $Q^2$-dependence of logarithmic QCD-motivated terms and power-like ones
may be studied separately 
(for a review, see the recent paper \cite{Beneke})
and references  therein).

In the present letter we sketch the results of our analysis \cite{KriKo}
(see also \cite{KriKo1})
at the next-to-leading order of perturbative QCD for
the most known DIS SF $F_2(x,Q^2)$ 
\footnote{Here $Q^2=-q^2$ and $x=Q^2/(2pq)$ are standard DIS variables,
where $q$ and $p$ are photon and hadron momentums, respectively.}
taking into account experimental data \cite{SLAC1}-\cite{BFP} of
SLAC, NM,  BCDMS and BFP Collaborations.
We
stress the power-like effects, so-called twist-4 (i.e.
$\sim 1/Q^2$)  contributions.
To our purposes we represent the SF $F_2(x,Q^2)$ as the contribution
of the leading twist part $F_2^{pQCD}(x,Q^2)$ 
described by perturbative QCD, 
when the target mass corrections are taken into account,
and the nonperturbative part 
%(``dynamical'' twist-four terms):
%\vskip -0.5cm
\begin{equation}
F_2(x,Q^2) 
%\equiv F_2^{full}(x,Q^2)
=F_2^{pQCD}(x,Q^2)\,
%\left(
\Bigl(
1+\frac{\tilde h_4(x)}{Q^2}
\Bigr),
\label{1}
\end{equation}
%\vskip -0.2cm
where $\tilde h_4(x)$ is magnitude of twist-4 terms.

Contrary to standard fits (see \cite{KriKo} and references therein)
when the direct numerical calculations based on 
Dokshitzer-Gribov-Lipatov-Altarelli-Parisi 
%(DGLAP) 
equation \cite{DGLAP} are used to evaluate structure functions, 
we use the exact solution of DGLAP equation
for the Mellin moments of
%$M_n^{pQCD}(Q^2)$ of
%$F_2^{full}(x,Q^2)$, $F_2^{pQCD}(x,Q^2)$ and
SF $F_2^{pQCD}(x,Q^2)$
and
the subsequent reproduction of $F_2^{pQCD}(x,Q^2)$
%$F_2^{full}(x,Q^2)$, $F_2^{pQCD}(x,Q^2)$  and/or $F_2^{tw2}(x,Q^2)$ 
at every needed $Q^2$-value with help of the Jacobi 
Polynomial expansion method \cite{Barker,Kri}
\footnote{We 
%would like to 
note here that there is similar method 
\cite{Ynd}, based on Bernstein polynomials. The method has been used 
in the analyses at the NLO level in \cite{KaKoYaF}
and at the NNLO level in \cite{SaYnd}.}
(see similar analyses at the NLO level 
\cite{Kri,Vovk}
and at the next-next-to-leading order (NNLO) level and above \cite{PKK}).

\vspace{-0.3cm}
\section{ Results of fits }

We have studied in
%demonstrated several steps of our study 
\cite{KriKo}
%of 
the $Q^2$-evolution of DIS structure function $F_2$ fitting all
modern fixed target experimental data 
at 
%Bjorken variable $x$ values: 
$x \geq 10^{-2}$ and $Q^2 \geq 1$ GeV$^2$. 
From the fits we have obtained the value of the normalization 
$\asMZ$
of QCD coupling constant. 

First of all, we have reanalyzed the BCDMS data 
cutting the range with large systematic errors. As it is possible to see
in \cite{KriKo}
%the Fig. 1, 
the value of $\asMZ$ rises strongly when
the cuts of systematics were incorporated. In another side, 
the value of $\asMZ$ does not dependent on the concrete type of the
cut within 
%in the range of 
modern statistical errors.

To verify the range of applicability of perturbative QCD,
we analyzed firstly the SLAC, BCDMS, NM and BFP
data without a contribution of twist-4 terms,
i.e. when $F_2 = F_2^{pQCD}$. We did several fits using the cut 
$Q^2 \geq Q^2_{cut}$ and increased the value $Q^2_{cut}$ step by step.
We observed  good agreement of the fits with the data when 
$Q^2_{cut} \geq 10 \div 15$ GeV$^2$ (see the Figs. 1 and 2).
Later we added the twist-4 corrections and fitted the data with the
standard cut $Q^2 \geq 1$ GeV$^2$.
We have found very good agreement with the data. Moreover, 
the predictions for $\asMZ$ in both above procedures 
are very similar (see the 
%Table 6 and 
Figs. 1 and 2).
%  Fig. 2,   Fig. 3
%\begin{figure}[htb]
\begin{figure}[t]
\vspace{-0.5cm}
%\begin{minipage}[t]{0.48\linewidth}
%\begin{center}
\includegraphics[width=2.7in]{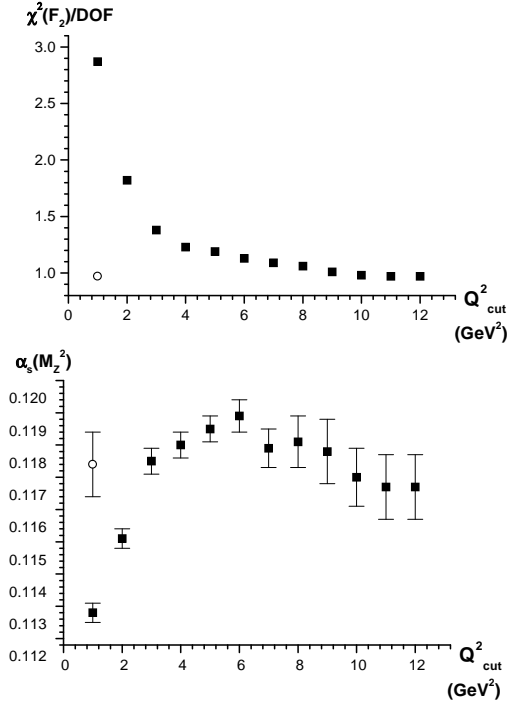} 
%\end{center}
\vspace{-0.6cm}
\caption{
The values of $\asMZ$ and $\chi^2$ at different $Q^2$-values of data cuts
in the fits based on nonsinglet evolution.
%regimes of fits. 
The black (white) 
points show the 
analyses of data without (with) twist-4 contributions.
Only statistical errors are shown.
\vskip -0.4cm
}
%\end{minipage}
 \label{fig:3}
\end{figure}

%\begin{figure}[htb]
\begin{figure}[t]
%\end{minipage}%
%\hspace{0.04\textwidth}%
\vspace{-0.5cm}
%\begin{minipage}[t]{0.48\linewidth}
%\begin{center}
\includegraphics[width=2.7in]{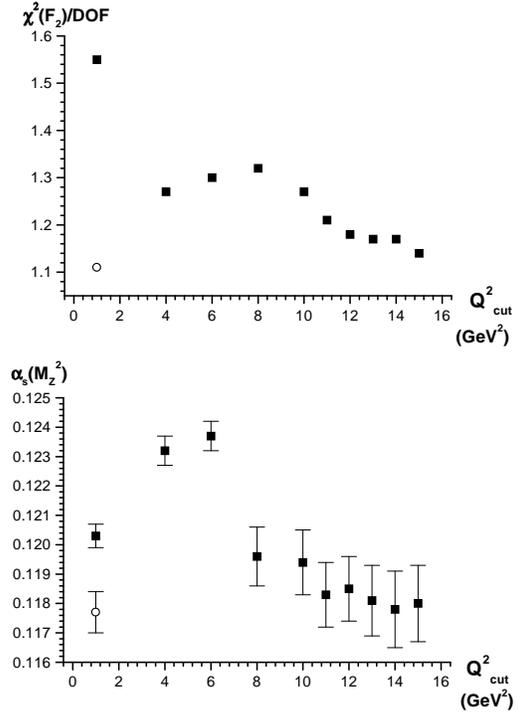} 
%\end{center}
\vspace{-0.7cm}
 \caption{
The values of $\asMZ$ and $\chi^2$ at different $Q^2$-values of data cutes
in the fits based on combine singlet and nonsinglet evolution. 
All other notes are as in Fig. 1.
\vskip -0.6cm
}\label{fig:4}
%\end{minipage}
\end{figure}

The results for  $\asMZ$ coincide for the both types of analyses:
%nonsinglet ones and singlet ones. They have the following form:
ones, based on
nonsinglet evolution, and ones, based on combined singlet and 
nonsinglet evolution.
%ones and singlet ones. 
They have the following form:
\bea
\as(M_Z^2) &=& 0.1177 \pm 0.0024 ~\mbox{(total)},
%0.0007 ~\mbox{(stat)}
%\pm 0.0021 ~\mbox{(syst)} \pm 0.0009 ~\mbox{(norm)}, 
\label{re2s}
\eea
%\end{itemize}
%\vskip -0.3cm
where the symbol ``total'' marks the total experimental error, which 
contains the sum of statistical error, systematic one and error of
normalization in quadratures.
%squred root 

We would like to note that the result (\ref{re2s}) is in
%we have 
good agreement with the analysis 
\cite{H1BCDMS} of
combined H1 and BCDMS data, which has been given by H1 Collaboration very 
recently. The result
%Our results 
for $\as(M_Z^2)$ is in good agreement also with 
the average value for coupling constant,
%for $\asMZ$, 
presented in 
%the recent studies (see, for example, \cite{Al2000}
%and references therein) and in
famous Altarelli and Bethke reviews \cite{Breview}.

\vskip 0.2cm
{\bf Acknowledgments.}~
%Authors
A.V.K. would like to express his sincerely thanks to the Organizing
  Committee of ACAT 2002
%the VIIIth International Workshop on advanced computing
%and analysis techniques in physics research (ACAT 2002) 
for the kind 
invitation.
%, the financial support at  such remarkable Conferences, and 
% for fruitful discussions. A.V.K. 
He was supported in part by 
%Alexander von Humboldt fellowship and 
INTAS  grant N366.

%\vspace{-0.3cm}

\end{document}